\documentclass[aps,preprint,amsmath,amssymb,superscriptaddress,nofootinbib]{revtex4}

\usepackage{graphicx}
\begin{document}

\title{Graviton production through photon-quark scattering at the LHC}

\author{\.{I}. \c{S}ahin}
\email[]{inancsahin@ankara.edu.tr}
 \affiliation{Department of
Physics, Faculty of Sciences, Ankara University, 06100 Tandogan,
Ankara, Turkey}

\author{M. K\"{o}ksal}
\email[]{mkoksal@cumhuriyet.edu.tr} \affiliation{Department of
Physics, Cumhuriyet University, 58140 Sivas, Turkey}

\author{S. C. \.{I}nan}
\email[]{sceminan@cumhuriyet.edu.tr}  \affiliation{Department of
Physics, Cumhuriyet University, 58140 Sivas, Turkey}

\author{A. A. Billur}
\email[]{abillur@cumhuriyet.edu.tr} \affiliation{Department of
Physics, Cumhuriyet University, 58140 Sivas, Turkey}

\author{B. \c{S}ahin}
\email[]{banusahin@ankara.edu.tr} \affiliation{Department of
Physics, Faculty of Sciences, Ankara University, 06100 Tandogan,
Ankara, Turkey}

\author{P. Tekta\c{s}}
\affiliation{Department of Physics, Faculty of Sciences, Ankara
University, 06100 Tandogan, Ankara, Turkey}

\author{E. Al{\i}c{\i}}
 \affiliation{Department of Physics, Bulent Ecevit University, 67100 Zonguldak, Turkey}

\author{R. Y{\i}ld{\i}r{\i}m}
  \affiliation{Department of
Physics, Cumhuriyet University, 58140 Sivas, Turkey}

\begin{abstract}
We have investigated real graviton emission in the ADD and RS model
of extra dimensions through the photoproduction process $pp\to
p\gamma p\to p G qX$ at the LHC. We have considered all
contributions from the subprocesses $\gamma q \to G q$, where
$q=u,d,c,s,b,\bar u,\bar d, \bar c, \bar s, \bar b$ quark. The
constraints on model parameters of the ADD and RS model of extra
dimensions have been calculated. During numerical calculations we
have taken account of 3, 4, 5 and 6 large extra dimensional
scenarios. The constraints on RS model parameters have been
calculated by considering $G \to \gamma \gamma, e \bar e, \mu \bar
\mu$ decay channels of the graviton.

\end{abstract}


\maketitle

\section{Introduction}

\begin{figure}
\includegraphics[scale=0.5]{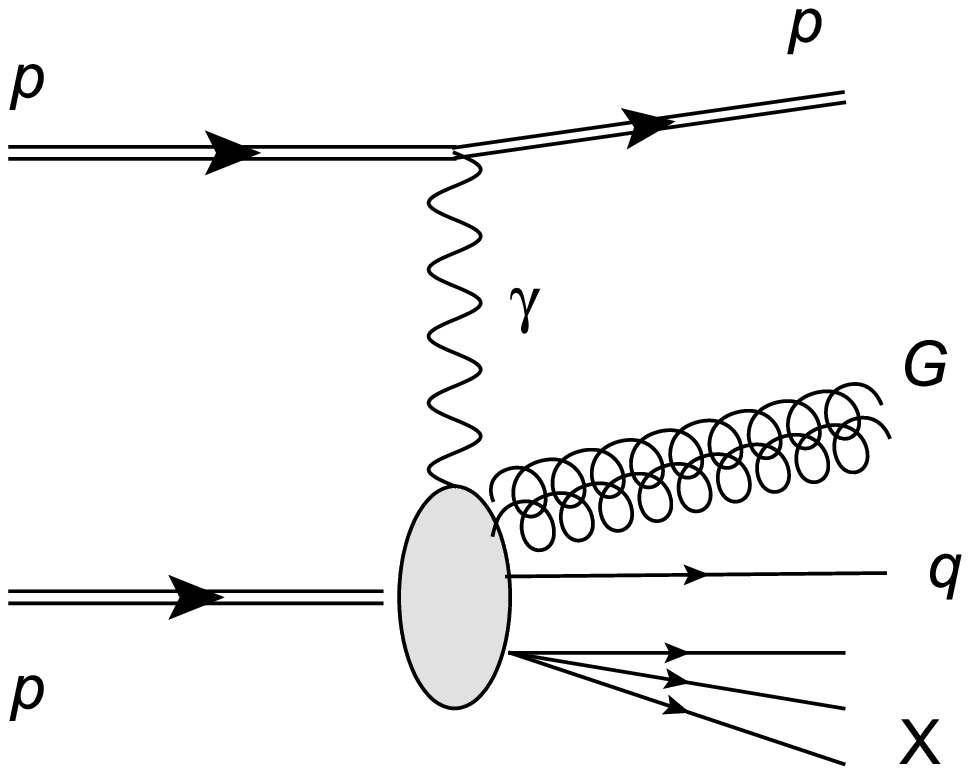}
\caption{The photoproduction process $pp\to p\gamma p\to p G qX$.
\label{fig1}}
\end{figure}

One of the leading aims of the Large Hadron Collider (LHC) is to
discover new physics beyond the Standard Model. In this respect,
extra dimensional models in particle physics have been drawing
attention for the past fifteen years. These models offer a solution
to the hierarchy problem, and provide possible candidates for dark
matter. Phenomenology of extra dimensional models at the LHC has
been widely studied in the literature as a possible new physics
candidate. These phenomenological studies generally involve usual
proton-proton deep inelastic scattering (DIS) processes where both
of the colliding protons dissociate into partons. Subprocesses of
quark-quark, gluon-gluon, and quark-gluon scattering have been
probed in detail. On the other hand, exclusive production and
semi-elastic processes have been much less studied in the
literature. In an exclusive production process both of the incoming
protons remains intact. They do not dissociate into partons. However
in a semi-elastic scattering process, one of the incoming proton
dissociate into partons but the other proton remains intact
\cite{rouby,Schul:2011xwa}. The exclusive and semi-elastic proton
processes are characterized by the interchange of photons or
pomerons. The former generally give larger cross sections since the
survival probability for processes involving a photon interchange is
larger than that for a pomeron interchange. The exclusive and
semi-elastic photon-mediated processes are sometimes called
two-photon and photoproduction processes respectively. Phenomenology
of extra dimensions has not been studied comprehensively in these
types of processes. Extra dimensional models have been examined via
the following two-photon processes: $pp\to p \gamma \gamma p \to p
\ell \bar \ell p$ \cite{sahinextradim1}, $pp\to p \gamma \gamma p
\to p \gamma \gamma p$ \cite{sahinextradim2,Sun:2014qba}, $pp\to p
\gamma \gamma p \to p \Phi p$ \cite{Goncalves:2010dw}, $pp\to p
\gamma \gamma p \to p t \bar t p$ \cite{inanbillurextradim} and
$pp\to p \gamma \gamma p \to p G p$ \cite{Inan:2012zz}. As far as we
know, except for our recent paper \cite{Sahin:2013qoa} phenomenology
of extra dimensional models has not been studied and model
parameters have not been constrained in any photoproduction process
at the LHC. In our recent paper we have analyzed the virtual effects
of Kaluza-Klein (KK) gravitons in $pp\to p\gamma p\to p\gamma qX$.
In this paper we will investigate real KK graviton emission through
the photoproduction process $pp\to p\gamma p\to p G qX$
(Fig.\ref{fig1}). $pp\to p\gamma p\to p G qX$ consists of the
following subprocesses:
\begin{eqnarray}
\label{subprocesses} &&\text{(i)}\;\;\gamma u \to G u
\;\;\;\;\;\;\;\;\;\;\;\;\text{(vi)}\;\;\gamma \bar u \to G \bar u
\nonumber
\\ &&\text{(ii)}\;\;\gamma d \to G d
\;\;\;\;\;\;\;\;\;\;\;\;\text{(vii)}\;\;\gamma \bar d \to G \bar d
\nonumber
\\&&\text{(iii)}\;\;\gamma c \to G c
\;\;\;\;\;\;\;\;\;\;\text{(viii)}\;\;\gamma \bar c \to G \bar c  \\
&&\text{(iv)}\;\;\gamma s \to G s
\;\;\;\;\;\;\;\;\;\;\;\text{(ix)}\;\;\gamma \bar s \to G \bar s
\nonumber \\&&\text{(v)}\;\;\gamma b \to G b
\;\;\;\;\;\;\;\;\;\;\;\;\;\text{(x)}\;\;\gamma \bar b \to G \bar b
\nonumber
\end{eqnarray}
Therefore, in order to obtain the cross section for the main process
$pp\to p\gamma p\to p G qX$ we have to consider all contributions
coming from subprocesses in (\ref{subprocesses}).

In a photoproduction process emitted photons from the proton ought
to carry a small amount of virtuality. (Otherwise, proton
dissociates after the emission.) Thus, the equivalent photon
approximation (EPA) can be successfully applied to a photoproduction
process. In EPA we employ the formalism of
\cite{budnev1975,baur2002,piotrzkowski2001}, and take account of the
electromagnetic form factors of the proton. Hence, EPA formula that
we have used is different from the pointlike electron or positron
case.

Exclusive two-photon and photoproduction processes can be
distinguished from fully inelastic processes by the virtue of the
following experimental signatures: After the elastic emission of a
photon, proton is scattered with a small angle and escapes detection
from the central detectors. This causes a missing energy signature
called {\it forward large-rapidity gap}, in the corresponding
forward region of the central detector
\cite{Albrow:2008az,rouby,Schul:2011xwa}. This technique was
successfully used at the Fermilab Tevatron by the CDF Collaboration.
Exclusive production of $\ell \bar \ell$, $\gamma\gamma$, $jj$ and
$J/\psi$ were observed experimentally
\cite{cdf1,cdf2,cdf3,cdf4,cdf5}. CMS Collaboration has also observed
exclusive production of $\ell \bar \ell$ and $W^+W^-$ pairs using
early LHC data at $\sqrt s$=7 TeV
\cite{Chatrchyan:2011ci,Chatrchyan:2012tv,Chatrchyan:2013foa}.
Another experimental signature can be implemented by forward proton
tagging. There are some proposals that aim to equip ATLAS and CMS
central detectors with very forward detectors (VFD) which can detect
intact scattered protons with a large pseudorapidity. Forward proton
tagging in VFD supports forward large-rapidity gap signatures
obtained from the central detectors. Operation of VFD in conjunction
with central detectors with a precise timing, can efficiently reduce
backgrounds from pile-up events
\cite{Albrow:2008pn,Tasevsky:2009zza,Albrow:2010yb,Tasevsky:2014cpa}.
It is argued that pile-up background grows rapidly with luminosity
and becomes important for high luminosity runs at the LHC.

Apart from extra dimensions other new physics scenarios have also
been studied via two-photon and photoproduction processes at the
LHC. Phenomenological studies involve new physics scenarios such as
supersymmetry, unparticles, technicolor, magnetic monopoles and
model independent analysis of anomalous interactions
\cite{pheno-1,pheno-2,pheno-3,pheno-4,pheno-5,pheno-6,pheno-7,pheno-8,pheno-9,pheno-10,
pheno-11,pheno-12,pheno-13,pheno-14,pheno-15,pheno-16,pheno-17,pheno-18,pheno-19,pheno-20,pheno-21,pheno-22,pheno-23,pheno-24,pheno-25,pheno-26,pheno-27,pheno-28,pheno-29,pheno-30,pheno-31,pheno-32,pheno-33}.

\section{Extra Dimensional Models and Cross Sections}
In this paper we will focus on two different extra dimensional
models, namely, Arkani-Hamed, Dimopoulos and Dvali (ADD) model of
large extra dimensions
\cite{ArkaniHamed:1998rs,ArkaniHamed:1998nn,Antoniadis:1998ig} and
Randall and Sundrum (RS) model of warped extra dimensions
\cite{Randall:1999ee}. The ADD model assumes a
$(4+\delta)$-dimensional spacetime where $\delta$ represents the
number of extra spatial dimensions. Extra dimensions are flat and
compactified in a volume $V_\delta$. For instance, we can consider a
toroidal compactification of volume $V_\delta=(2\pi R)^\delta$.
Here, $R$ represents the radii of extra dimensions. In the ADD
model, extra dimensions can be as large as approximately 0.1 mm. For
this reason ADD model is sometimes called the large extra
dimensional model. In principle, the number of extra dimensions can
be $\delta\geq1$. But astronomical observations on Newton's
gravitation law rule out $\delta=1$ case. There are also constraints
from table top experiments and astrophysical observations. These
constraints are stringent for $\delta=2$. Hence, we will consider
the case in which $\delta\geq3$. The ADD model solves the hierarchy
problem by introducing a new mass scale $M_D$ called the fundamental
scale which is at the order of electroweak scale. $M_D$ is the mass
scale of the $(4+\delta)$-dimensional theory. Thus, it is argued by
the model that the true mass scale of the theory is $M_D$, not the
Planck scale $M_{Pl}$. On the other hand, an observer confined in a
4-dimensional spacetime measures its value as $M_{Pl}$.

The original version of the RS model assumes the existence of only
one extra spatial dimension and two 3-branes (4-dimensional
spacetimes) located at boundaries of the extra dimensional
coordinate $y$. The metric is given by \cite{Randall:1999ee}
\begin{eqnarray}
\label{RSmetric}
ds^{2}=e^{-2k|y|}\eta_{\mu\nu}dx^{\mu}dx^{\nu}-dy^{2},
\end{eqnarray}
where $k$ represents a constant of the order of the Planck scale.
The metric (\ref{RSmetric}) describes a 5-dimensional anti-de Sitter
space with a cosmological constant $\Lambda$. Extra dimensional
coordinate $y$ can be parametrized by an angular coordinate $\phi$
via $y=r_c\phi$. Here, $r_c$ represents the compactification radius
of the extra dimension. The angular coordinate varies in the range
$0\leq|\phi|\leq\pi$. It is assumed that we are living in the
3-brane located at the boundary point $\phi=\pi$. It is deduced from
the model that any mass scale $m$ observed in $\phi=\pi$ brane is
generated from the fundamental mass scale $m_0$ which is at the
order of $M_{Pl}$. The relation between these two scales is given
by: $m=m_0e^{-kr_c\pi}$. Thus, the hierarchy problem is eliminated.

Both the ADD and RS model of extra dimensions involve massive
gravitons and gravitational scalars which are possible dark matter
candidates. On the other hand, mass scales for these new particles
are very different in the ADD and RS models. This leads to different
phenomenologies. In the ADD model of extra dimensions, mass
difference between consecutive graviton excited states is given
approximately by the formula \cite{Giudice:1998ck,Han:1998sg}:
\begin{eqnarray}
\Delta m\approx \frac{M_D^{\frac{2+\delta}{\delta}}}{\bar
{M_{Pl}}^\frac{2}{\delta}}
\end{eqnarray}
We see from this formula that mass splitting is quite narrow. For
instance, if we choose $M_D=1$ TeV and $\delta=4$ then $\Delta
m\approx 50$ KeV. Therefore, at collider energies KK graviton
production processes involve a summation over huge number of
graviton final states. This summation can be approximated to an
integral and the cross section for the inclusive production process
can be written as \cite{Giudice:1998ck}
\begin{eqnarray}
\label{ADDtotalcross} \sigma=\int\int \left( \frac{d^2\sigma}{dmdt}
\right) dm dt
\end{eqnarray}
\begin{eqnarray}
\label{ADDdiffcross}
\frac{d^2\sigma}{dmdt}=\frac{2\pi^{\delta/2}}{\Gamma(\delta/2)}\frac{\bar
{M_{Pl}}^2}{{M_D}^{2+\delta}}m^{\delta-1}\frac{d\sigma_m}{dt}
\end{eqnarray}
where $\frac{d\sigma_m}{dt}$ is the cross section for producing a
single KK graviton state of mass $m$. The differential cross section
$\frac{d\sigma_m}{dt}$ can be obtained in conventional way from the
scattering amplitude. Since the masses of KK states are integrated,
total cross section in Eq. (\ref{ADDtotalcross}) depends on two
model parameters $M_D$ and $\delta$.

In the RS model case, the spectrum of KK graviton masses is given by
\cite{Davoudiasl:1999jd}
\begin{eqnarray}
m_{n}=x_{n}ke^{-kr_{c}\pi}=x_{n}\beta\Lambda_{\pi},\;\;\;\;\beta=k/\bar{M}_{Pl}
\end{eqnarray}
where $x_{n}$ are the roots of the Bessel function $J_1$, i.e.,
$J_{1}(x_{n})=0$. We can deduce from this formula that mass
splitting between consecutive KK graviton states is considerably
large, $\sim {\cal O}{(TeV)}$. Hence it is assumed that only the
first KK graviton state can be observed. We will represent the mass
of this first KK graviton state by $m_G$. Masses of other KK states
are proportional to $m_G$: $m_n=\frac{x_n}{x_1}m_G$. In the RS model
we have two independent model parameters. We prefer to choose $m_G$
and $\beta$ as independent parameters.

The process $\gamma q \to G q$ is represented by the Feynman
diagrams in Fig.\ref{fig2}. We do not give the Feynman rules for KK
gravitons. One can find the Feynman rules in Refs.
\cite{Giudice:1998ck,Han:1998sg}. Analytical expressions for the
scattering amplitude are given in the Appendix. As we have mentioned
in the introduction we consider all possible initial quark flavors.
Thus, in order to obtain the total cross section for the process
$pp\to p\gamma p\to p G qX$ we integrate the subprocess cross
sections over the photon and quark distributions and sum all
contributions from different subprocesses:
\begin{eqnarray}
\label{mainprocess}
 \sigma\left(pp\to p\gamma p\to p G qX \right)=\sum_q\int_{\xi_{min}}^{\xi_{max}} {dx_1 }\int_{0}^{1}
{dx_2}\left(\frac{dN_\gamma}{dx_1}\right)\left(\frac{dN_q}{dx_2}\right)
{\sigma}({\gamma q\to G q})
\end{eqnarray}
Here, $x_1$ represents the energy ratio between the equivalent
photon and incoming proton and $x_2$ is the fraction of the proton's
momentum carried by the struck quark. $\frac{dN_\gamma}{dx_1}$ and
$\frac{dN_q}{dx_2}$ are the equivalent photon and quark distribution
functions. The analytical expression for $\frac{dN_\gamma}{dx_1}$
can be found in the literature, for example in
\cite{budnev1975,baur2002,piotrzkowski2001,Sahin:2013qoa} or
\cite{pheno-7}. $\frac{dN_q}{dx_2}$ can be evaluated numerically. In
this paper we have used parton distribution functions of Martin {\it
et al.} \cite{Martin:2009iq}.

\section{Numerical Results}
During numerical calculations we assume the existence of VFD
proposed by Refs. \cite{Albrow:2008pn,royon2007,avati2006,pheno-7}.
In Refs. \cite{Albrow:2008pn,royon2007}, it was proposed to locate
VFD at 220 m and 420 m distances away from the ATLAS interaction
point. These detectors can detect forward protons within the
acceptance region $0.0015<\xi<0.15$. Here,
$\xi\equiv(|\vec{p}|-|\vec{p}^{\,\,\prime}|)/|\vec{p}|$ where $\vec
p$ represents the initial proton's momentum and
$\vec{p}^{\,\,\prime}$ represents forward proton's momentum after
scattering. According to another scenario VFD can detect forward
protons within $0.0015<\xi<0.5$ \cite{avati2006,pheno-7}. This
scenario is based on VFD located at 420 m distance from the CMS
interaction point and TOTEM detectors at 147 m and 220 m.

For all numerical results presented in this paper, the
center-of-mass energy of the proton-proton system is taken to be
$\sqrt s=14$ TeV and the virtuality of the DIS is taken to be
$Q^2={(5M_Z)}^2\approx (456\; GeV)^2$. Here, $M_Z$ is the mass of
the Z boson and it represents only a scale which is roughly at the
order of Standard Model energies. This virtuality value is
reasonable since the average value for the square of the momentum
transferred to the proton is approximately at that order. Moreover,
at high energies parton distribution functions do not depend
significantly on virtuality. Thus, it is reasonable to use a fix
virtuality value for the DIS.

In Fig.\ref{fig3} and Fig.\ref{fig4} we present ADD model cross
section of the process $pp\to p\gamma p\to p G qX$ as a function of
the fundamental scale $M_D$ for VFD acceptances of $0.0015<\xi<0.5$
and $0.0015<\xi<0.15$ respectively. We consider the cases in which
the number of extra spatial dimensions are $\delta=3,4,5,6$. We see
from these figures that cross section decreases with increasing
$\delta$. Moreover, cross sections for different number of extra
dimensions significantly deviate from each other for large $M_D$
values. These behaviors are obvious from differential cross section
formula in Eq. (\ref{ADDdiffcross}). As we have discussed in the
previous section, ADD graviton final states consist of huge number
of KK states arranged densely with increasing mass. This sequence of
states is sometimes called a KK tower. A KK tower is not detected
directly in the detectors. Instead, its presence is inferred from
missing energy signal. Missing signal for a KK tower exhibits a
peculiarity which is very different compared with any other new
physics processes \cite{Giudice:1998ck}. The peculiarity originates
from continuous mass distribution of graviton states in a KK tower.
Different from other new physics or Standard Model processes, we do
not have a fix final state mass. But the cross section is integrated
over the mass of final state gravitons (see Eq.
(\ref{ADDtotalcross})). Thus, the momentum carried by a KK tower can
take values which are kinematically forbidden in the fix mass case.
This peculiar behavior is reflected in missing transverse momentum
and invariant mass distributions and useful when we want to discern
graviton production from other new physics processes and Standard
Model backgrounds. In Fig.\ref{fig5}-Fig.\ref{fig8} we present
missing $p_T$ ($p_T$ of the KK tower) dependence of the cross
section of the process $pp\to p\gamma p\to p G qX$ for various
number of extra dimensions $\delta$. In these figures VFD acceptance
is taken to be $0.0015<\xi<0.5$ and $M_D$=5 TeV. $p_T$ distributions
for $0.0015<\xi<0.15$ case exhibit similar behaviors.

The main background to our process is $p p\to p\gamma p\to p \nu
\bar \nu q X$. This background process consists of the following
subprocesses: $\gamma q \to \nu \bar \nu q$ where $q=u,d,c,s,b,\bar
u,\bar d, \bar c, \bar s, \bar b$ quark and
$\nu=\nu_e,\nu_\mu,\nu_\tau$. Since the final state neutrinos are
not detected in the detectors they may generate a missing signal
similar to the one coming from graviton production. But as we have
discussed before, the missing signal associated with final state
neutrino pair exhibits a different behavior and can be discerned
from the missing signal associated with a KK tower. In
Fig.\ref{fig9} and Fig.\ref{fig10} we plot missing transverse
momentum and missing invariant mass dependence of the cross section
both for our process and the Standard Model background. In these
figures cross sections have been calculated by considering all
contributions from possible subprocesses. As can be seen from
Fig.\ref{fig9} and Fig.\ref{fig10} signal and background cross
sections are well separated from each other. We observe from
Fig.\ref{fig9} that differential cross section for the background
rapidly decreases as $p_T$ increases and it is suppressed for
$p_T>300$ GeV. Graviton production cross section also decreases with
increasing $p_T$. However, the decrease is approximately linear up
to $p_T\approx 3000$ GeV (not shown in the figure). The missing
invariant mass $M_{Inv}=\sqrt{E^2_{miss}-|\vec P_{miss}|^2}$
dependence of the cross section exhibit similar behaviors. We see
from Fig.\ref{fig10} that background cross section rapidly decreases
after the peak point around $M_{Inv}\approx M_Z$ and it is
suppressed for $M_{Inv}>600$ GeV. On the other hand, graviton
production cross section does not change considerably for $600\;
\text{GeV} <M_{Inv}<5000\; \text{GeV}$ (not shown in the figure).
Hence, the background can be discerned from the ADD signal and can
be eliminated by imposing a cut on the missing invariant mass or
missing transverse momentum.

In the RS model, we aim to observe first KK graviton state with mass
$m_G$. Similar to techniques used to detect a massive particle such
as $W,Z$ boson or top quark, an RS graviton can be detected from its
decay products. Angular distribution of the RS graviton for the
process $pp\to p\gamma p\to p G qX$  is given in Fig.\ref{fig11}. We
consider VFD acceptance of $0.0015<\xi<0.5$ but we expect a similar
behavior for $0.0015<\xi<0.15$ case. In the figure, we show both the
the cross section including the sum of all contributions from
subprocesses in (\ref{subprocesses}) and cross sections including
individual contributions from each subprocess.

We have obtained 95\% confidence level (CL) bounds on the model
parameters of the ADD and RS models. Since the Standard Model
contribution to the process is absent and the backgrounds can be
eliminated it is appropriate to employ a statistical analysis using
a Poisson distribution. In the ADD model case, the expected number
of events is calculated from the formula: $N=S\times E\times
\sigma(pp\to p\gamma p\to p G qX)\times L_{int}$. In this formula,
$S$ represents the survival probability factor, $E$ represents the
jet reconstruction efficiency and $L_{int}$ represents the
integrated luminosity. We take into account a survival probability
factor of $S=0.7$ and jet reconstruction efficiency of $E=0.6$. We
also place a pseudorapidity cut of $|\eta|<2.5$ for final quarks and
antiquarks from subprocesses in (\ref{subprocesses}). In Tables
\ref{tab1} and \ref{tab2} we present 95\% CL bounds on the
fundamental scale for various values of $L_{int}$ and $\delta$
without imposing a cut on the missing invariant mass or missing
transverse momentum. In Tables \ref{tab3} and \ref{tab4} we present
similar bounds but in this case we impose a missing invariant mass
cut of $M_{Inv}>600$ GeV. As we have discussed, this cut effectively
eliminates the background contribution. We observe from Tables
\ref{tab1} - \ref{tab4} that the effect of the cut on the
sensitivity bounds is minor. The bounds on $M_D$ are slightly
spoiled. For instance, when we compare the bounds in Tables
\ref{tab1} and \ref{tab3} we see that the percentage differences
between bounds with and without a cut do not exceed $4\%$, $1\%$,
$0.5\%$ and $0.3\%$ for $\delta=3,4,5$ and 6 respectively. In the RS
model, bounds are calculated in the plane of $\beta$ versus $m_G$.
Since an RS graviton can be detected from its decay products, the
expected number of events is given by $N=S\times E\times
\sigma(pp\to p\gamma p\to p G qX)\times L_{int}\times BR$, where
$BR$ represents the branching ratio for the graviton. We place a
pseudorapidity cut of $|\eta|<2.5$ for final quarks, antiquarks and
also the graviton. The limits have been calculated by considering $G
\to \gamma \gamma, e \bar e, \mu \bar \mu$ decay channels of the RS
graviton with a total branching ratio of $8\%$
\cite{Allanach:2002gn}. In Fig.\ref{fig12} we present the excluded
regions in the $\beta$ versus $m_G$ parameter plane for the above
branching ratio value. The Standard Model processes $pp\to p\gamma
p\to p\; (\gamma \gamma, e \bar e, \mu \bar \mu) qX$ give rise to
the same final states. Determination of an on-shell graviton with
mass ${\cal O}{(TeV)}$ requires an invariant mass measurement of the
final state charged lepton and photon pairs. Therefore, we should
impose a cut of $M_{\gamma \gamma,e\bar e,\mu \bar \mu}\approx
m_G\approx {\cal O}{(TeV)}$ on the invariant mass of final leptons
and photons. This cut reduces the effect of background processes
drastically. To be precise, the sum of all Standard Model
contributions, i.e., sum of the cross sections for the processes
$pp\to p\gamma p\to p\; k \bar k qX$ where $k=\gamma, e,\mu$ and
$q=u,d,c,s,b,\bar u,\bar d, \bar c, \bar s, \bar b$ quark, provides
a huge cross section of $1.3\times 10^{3}$ pb. But if we demand that
the invariant mass of final state lepton and photon pairs are in the
interval $990\; \text{GeV}<M_{\gamma \gamma,e\bar e,\mu \bar
\mu}<1010\; \text{GeV}$, then the total Standard Model contribution
becomes only $2.4\times 10^{-6}$ pb. Thus, the background
contribution is reduced by approximately a factor of $10^{9}$. The
number of Standard Model events is smaller than 1 and therefore the
background contributions can be ignored.

\section{Conclusions}

Processes involving real graviton final states provide a direct
signal for TeV-scale gravity. In the ADD model of extra dimensions,
KK gravitons behave like non-interacting stable particles and their
presence are inferred from missing energy signal
\cite{Giudice:1998ck}. But as we have discussed in the previous
section, missing signal associated with ADD gravitons exhibits a
peculiar behavior which is very different compared with any other
new physics models and Standard Model backgrounds. This peculiar
behavior allows us to isolate signals coming from graviton
production. In the RS model of extra dimensions, KK gravitons can be
detected via their decay products. The angular distribution of its
decay products can be used to determine the spin of the graviton
\cite{Allanach:2002gn}. In case a particle having a spin of 2 and a
mass of ${\cal O}{(TeV)}$ is detected, this will be a distinctive
signature for the model. Many new physics scenarios can be
eliminated by means of this signature. As we have shown, invariant
mass cut on the decay products of the graviton effectively
eliminates the background contributions. Therefore, graviton
production processes provide important clues for the models of
TeV-scale gravity. These type of processes give us the opportunity
to isolate and discern the model which is much more difficult for
processes in which virtual effects of KK gravitons are considered.

Virtual effects of KK gravitons were examined in the following
two-photon and photoproduction processes at the LHC: $pp\to p \gamma
\gamma p \to p \ell \bar \ell p$ \cite{sahinextradim1}, $pp\to p
\gamma \gamma p \to p \gamma \gamma p$
\cite{sahinextradim2,Sun:2014qba}, $pp\to p \gamma \gamma p \to p t
\bar t p$ \cite{inanbillurextradim} and $pp\to p\gamma p\to p\gamma
qX$ \cite{Sahin:2013qoa}. A comparison of our limits with the limits
obtained in these processes is difficult in the case of the ADD
model. It is because in these papers authors used the cutoff
procedure of Giudice {\it et al.} \cite{Giudice:1998ck} in the
graviton propagator that the cross section depends only on the
cutoff scale. It is independent of the number of extra dimensions
$\delta$. On the other hand, if we assume that the convention of
Giudice {\it et al.} corresponds to $\delta=4$ case \footnote{Han
{\it et al.} \cite{Han:1998sg} employed a different cutoff
procedure. According to the convention of Han {\it et al.} the
summation of KK states in the propagator is approximated to an
expression that depends on $\delta$. Conventions of Han {\it et al.}
and Giudice {\it et al.}  coincide for $\delta=4$.} then our present
limits on the fundamental scale are better than the limits obtained
in two-photon processes
\cite{sahinextradim1,sahinextradim2,Sun:2014qba,inanbillurextradim}
but approximately at the same order with the limits of our recent
paper  \cite{Sahin:2013qoa}. When we compare our RS limits with the
limits from processes mentioned above we see that our limits in the
plane of $\beta$ versus $m_G$ are stronger than the limits obtained
in \cite{sahinextradim1,inanbillurextradim} but approximately at the
same order with the limits of \cite{sahinextradim2,Sahin:2013qoa}.
To be precise, our limits for Atlas VFD scenario are little better
but our limits for CMS-TOTEM VFD scenario are little worse than the
corresponding limits obtained in \cite{Sahin:2013qoa}. In Ref.
\cite{sahinextradim2} only CMS-TOTEM VFD scenario was considered for
the RS model case. Thus, a comparison for Atlas VFD scenario is
impossible. Real graviton and radion final states were investigated
in the following two-photon processes: $pp\to p \gamma \gamma p \to
p Gp$ \cite{Inan:2012zz} and $pp\to p \gamma \gamma p \to p \Phi p$
\cite{Goncalves:2010dw}. In Ref. \cite{Goncalves:2010dw} authors
considered only the RS model case and it seems they did not obtain
the limits on RS model parameters. In Ref. \cite{Inan:2012zz} the
full decay channel of the RS graviton is considered i.e.,
$BR=100\%$. Hence, it is very difficult to compare our RS limits
with the corresponding limits of \cite{Inan:2012zz}. On the other
hand, our ADD limits are a percentage from $250\%$ to $500\%$ better
than the limits obtained in \cite{Inan:2012zz} depending on the
luminosity, VFD acceptance and $\delta$.

The current best experimental bounds on the ADD and RS model
parameters are provided by the ATLAS and CMS collaborations at the
CERN LHC
\cite{Marionneau:2013fna,ADD1,ADD2,ADD3,Chatrchyan:2012it,Aad:2012hf,Aad:2012bsa,Aad:2012cy,Aad:2014wca,Khachatryan:2014rra}.
The most stringent bounds on the ADD model parameters to date have
been obtained in proton-proton collisions with a center-of-mass
energy of $\sqrt s=8$ TeV and an integrated luminosity of
approximately $L_{int}=20fb^{-1}$
\cite{Marionneau:2013fna,Aad:2014wca,Khachatryan:2014rra}. When we
compare our ADD limits with the most stringent experimental limits
obtained at the LHC we see that our limits for a VFD acceptance of
$0.0015<\xi<0.5$ and an integrated luminosity of
$L_{int}=200fb^{-1}$ are a percentage from $125\%$ to $150\%$ better
than these experimental bounds depending on the value of $\delta$.
Experimental bounds on the RS model parameters have been obtained by
considering graviton resonances decaying to $e^-e^+$, $\mu^-\mu^+$
and $\gamma\gamma$ final states
\cite{Chatrchyan:2012it,Aad:2012hf,Aad:2012cy}. These experimental
bounds are stronger than the bounds that we have obtained. Our RS
limits for similar final states (Fig.\ref{fig12}) are approximately
$50\%$ worse than these experimental limits. In conclusion, the
photoproduction process $pp\to p\gamma p\to p G qX$ at the LHC
possesses a remarkable potential in searching for the extra
dimensional scenario proposed by the ADD model. On the contrary, its
potential is low in probing warped extra dimensional model of RS.
Nevertheless, $pp\to p\gamma p\to p G qX$ process can also be used
in combination with other processes in order to verify the
universality feature of the graviton's coupling.

\begin{acknowledgments}
This work has been supported by the Scientific and Technological
Research Council of Turkey (T\"{U}B\.{I}TAK) in the framework of
the project no: 112T085.
\end{acknowledgments}

\appendix*
\section{Analytical expressions for the scattering amplitude}
\begin{figure}
\includegraphics[scale=1]{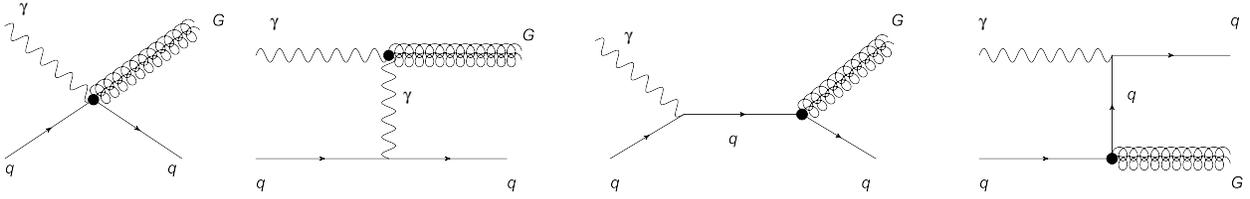}
\caption{Tree-level Feynman diagrams for the subprocess $\gamma q
\to G q$ ($q=u,d,c,s,b,\bar u,\bar d, \bar c, \bar s, \bar b$).
\label{fig2}}
\end{figure}

\begin{figure}
\includegraphics[scale=1]{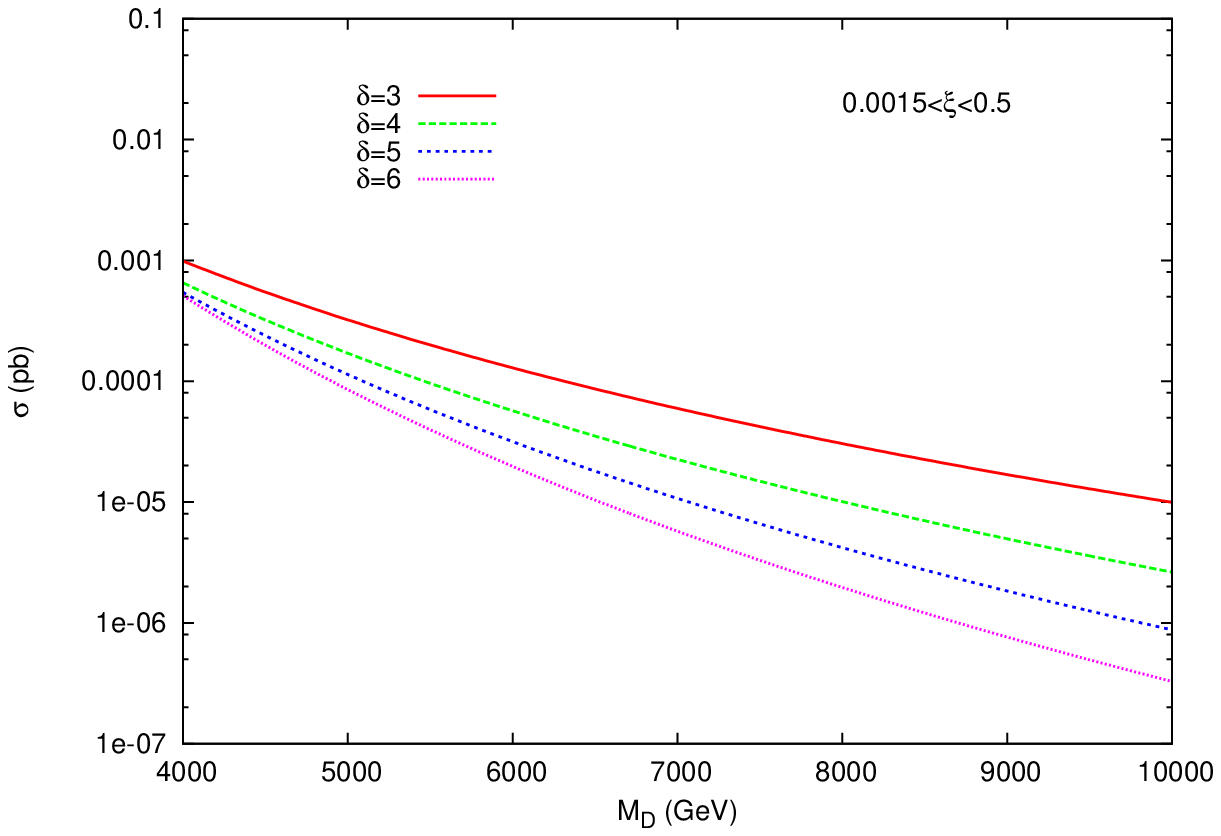}
\caption{The total cross section of the process $pp\to p\gamma p\to
p G qX$ as a function of the fundamental scale $M_D$ for various
number of extra dimensions $\delta$. The forward detector acceptance
is chosen to be $0.0015<\xi<0.5$. \label{fig3}}
\end{figure}

\begin{figure}
\includegraphics[scale=1]{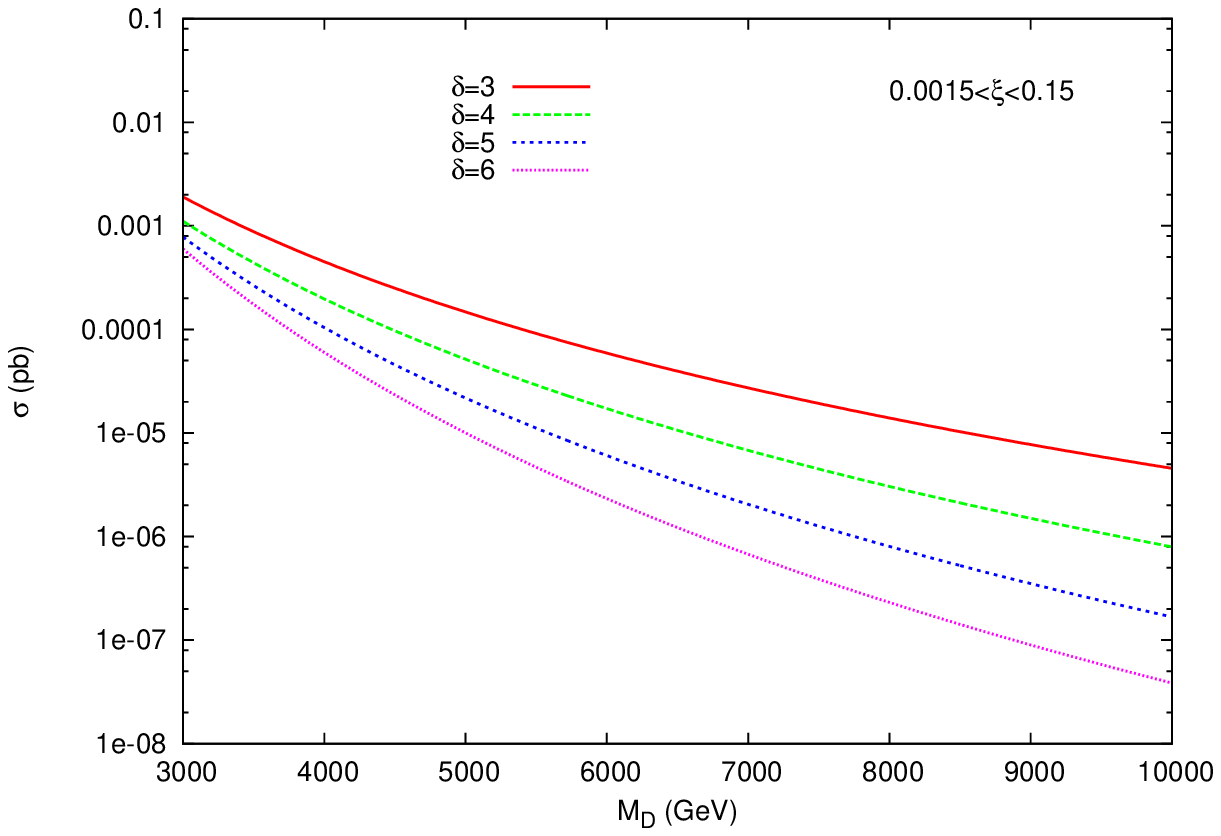}
\caption{The total cross section of the process $pp\to p\gamma p\to
p G qX$ as a function of the fundamental scale $M_D$ for various
number of extra dimensions $\delta$. The forward detector acceptance
is chosen to be $0.0015<\xi<0.15$. \label{fig4}}
\end{figure}

\begin{figure}
\includegraphics[scale=1]{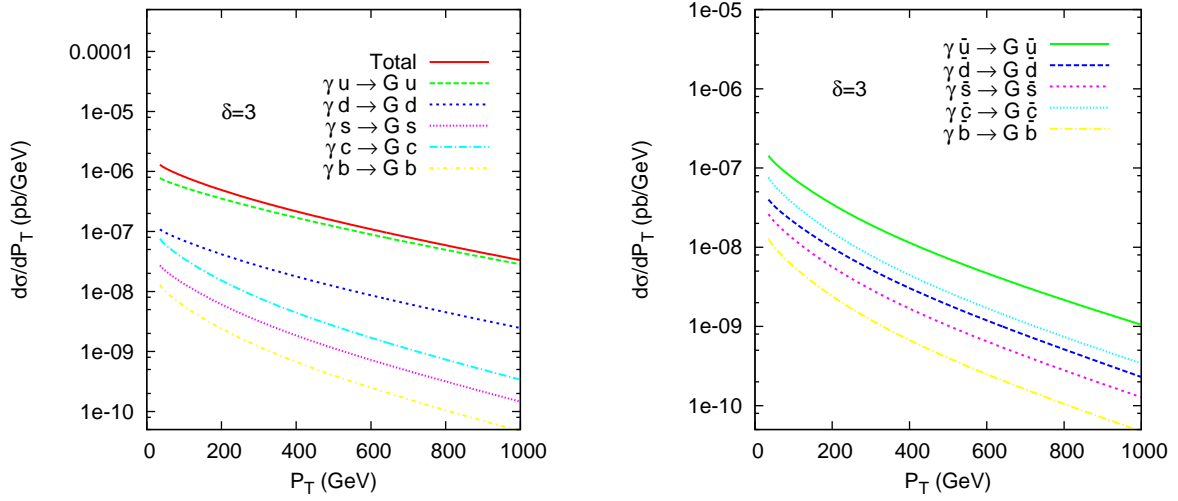}
\caption{The differential cross section of the process $pp\to
p\gamma p\to p G qX$ as a function of the transverse momentum of the
KK tower. The number of extra dimensions is chosen to be $\delta=3$
and $M_D=5$ TeV. In the left panel we present total cross section
(solid line) which represents the sum of all contributions from
subprocesses in (\ref{subprocesses}) and cross sections including
individual contributions from each subprocess with quarks. In the
right panel we present similar plots but for subprocesses with
antiquarks.\label{fig5}}
\end{figure}

\begin{figure}
\includegraphics[scale=1]{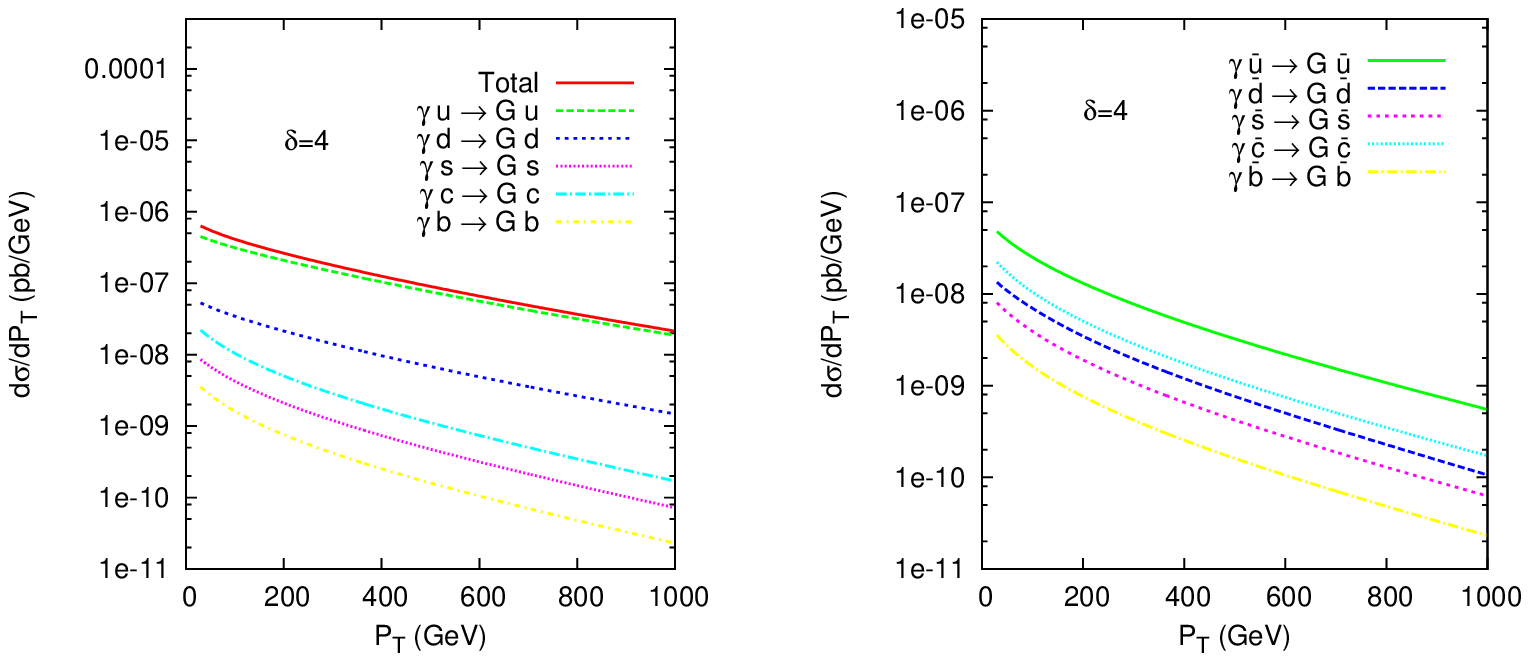}
\caption{The same as figure \ref{fig5} but for
$\delta=4$.\label{fig6}}
\end{figure}

\begin{figure}
\includegraphics[scale=1]{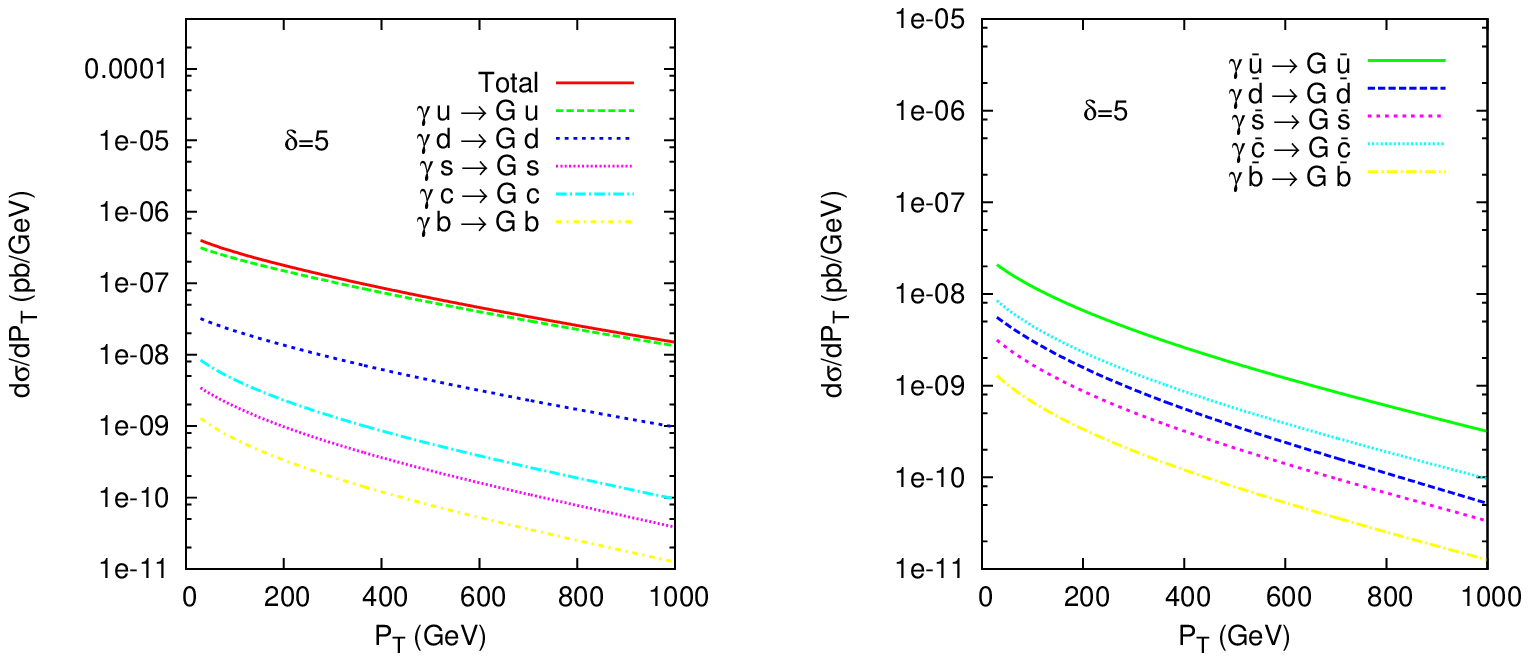}
\caption{The same as figure \ref{fig5} but for
$\delta=5$.\label{fig7}}
\end{figure}

\begin{figure}
\includegraphics[scale=1]{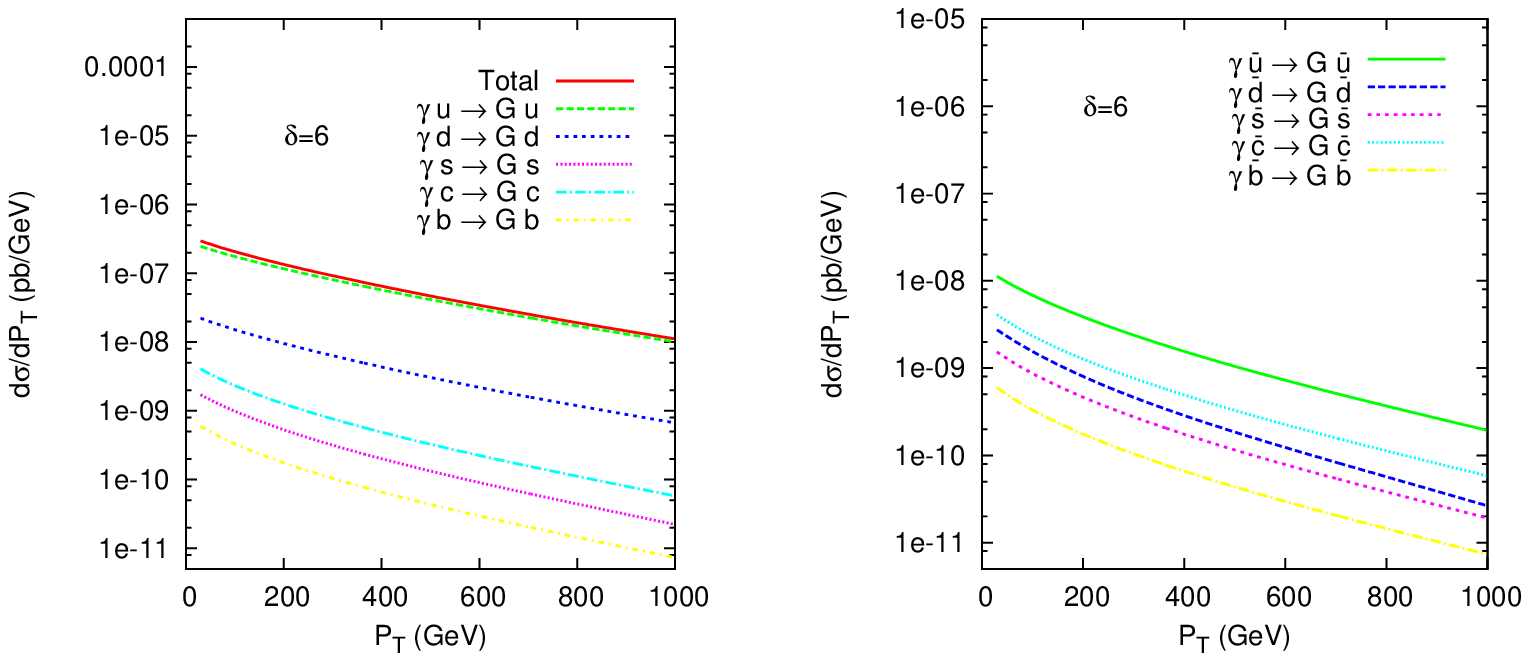}
\caption{The same as figure \ref{fig5} but for
$\delta=6$.\label{fig8}}
\end{figure}

\begin{figure}
\includegraphics[scale=1]{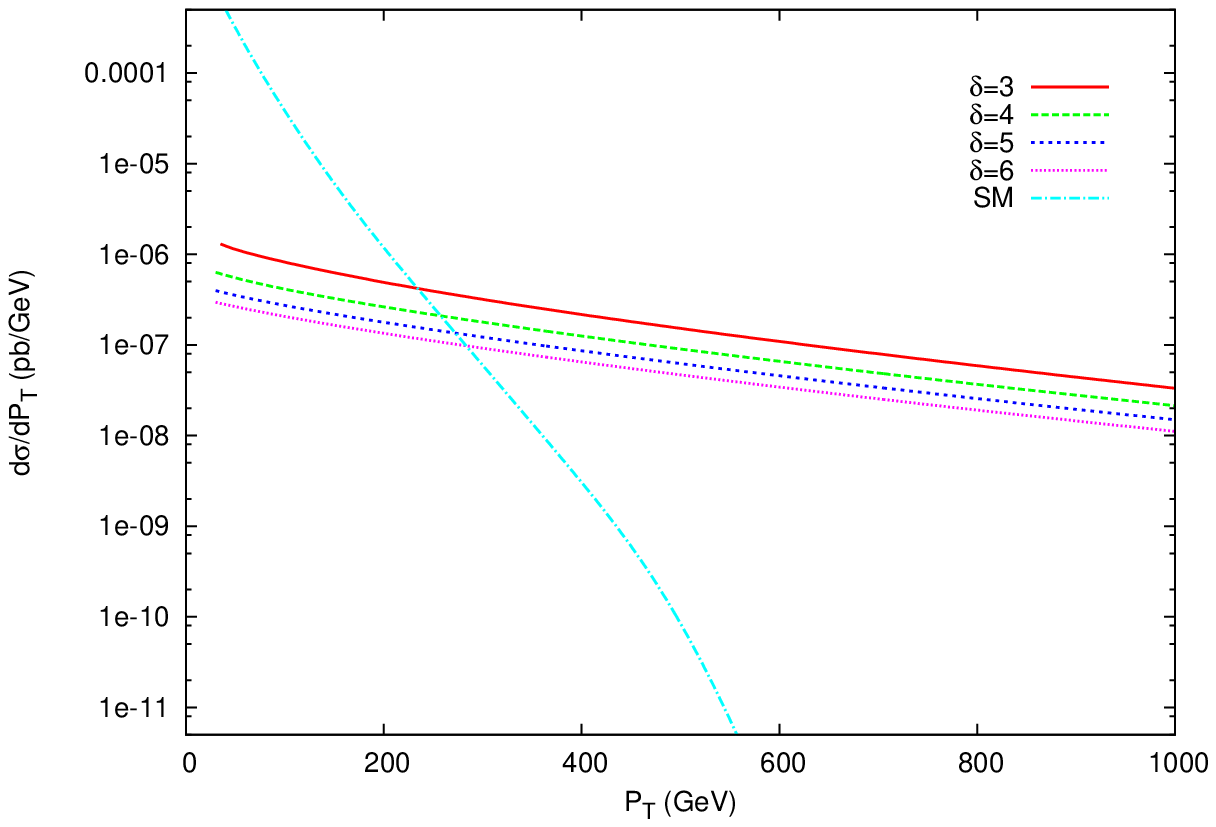}
\caption{The missing transverse momentum dependence of the cross
section for the process $pp\to p\gamma p\to p G qX$ and Standard
Model (SM) background in the center-of-momentum frame of the
proton-proton system. VFD acceptance is taken to be $0.0015<\xi<0.5$
and $M_D$=5 TeV.\label{fig9}}
\end{figure}

\begin{figure}
\includegraphics[scale=1]{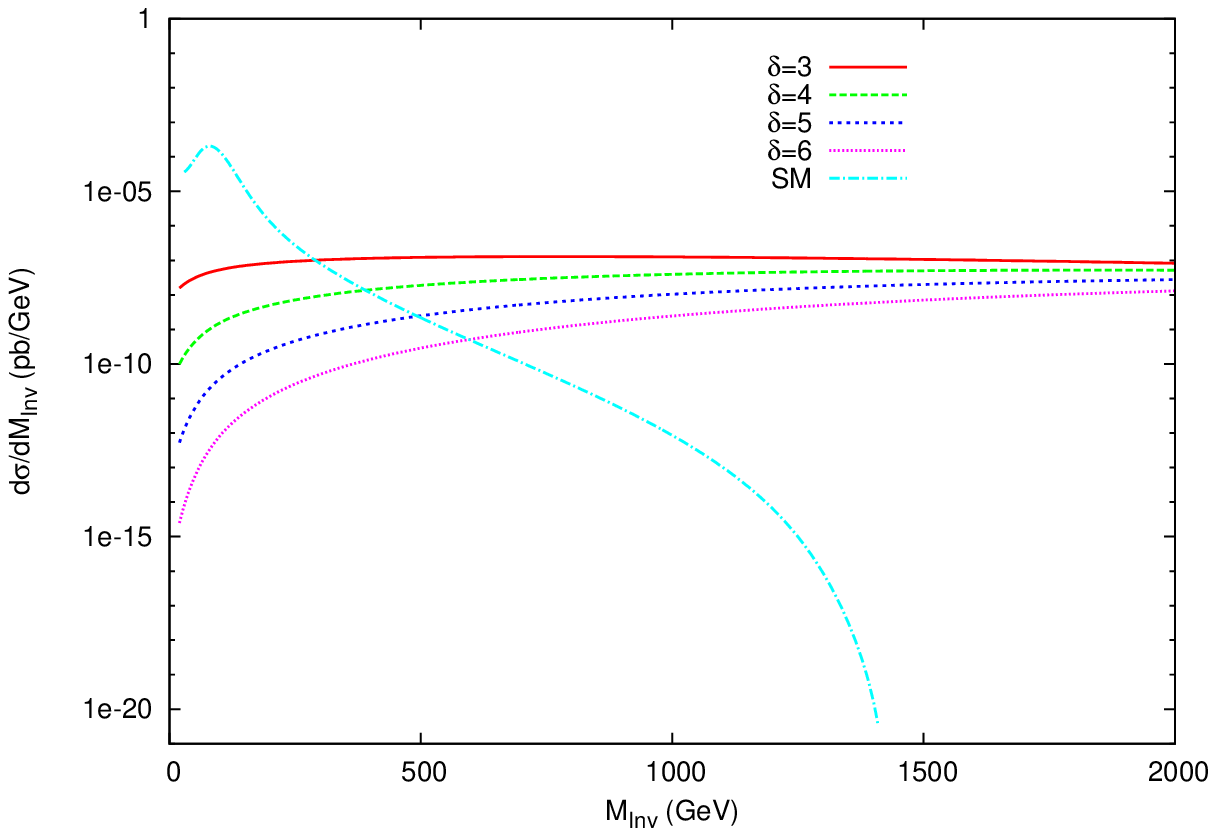}
\caption{The missing invariant mass dependence of the cross section
for the process $pp\to p\gamma p\to p G qX$ and Standard Model (SM)
background in the center-of-momentum frame of the proton-proton
system. VFD acceptance is taken to be $0.0015<\xi<0.5$ and $M_D$=5
TeV.\label{fig10}}
\end{figure}

\begin{figure}
\includegraphics[scale=1]{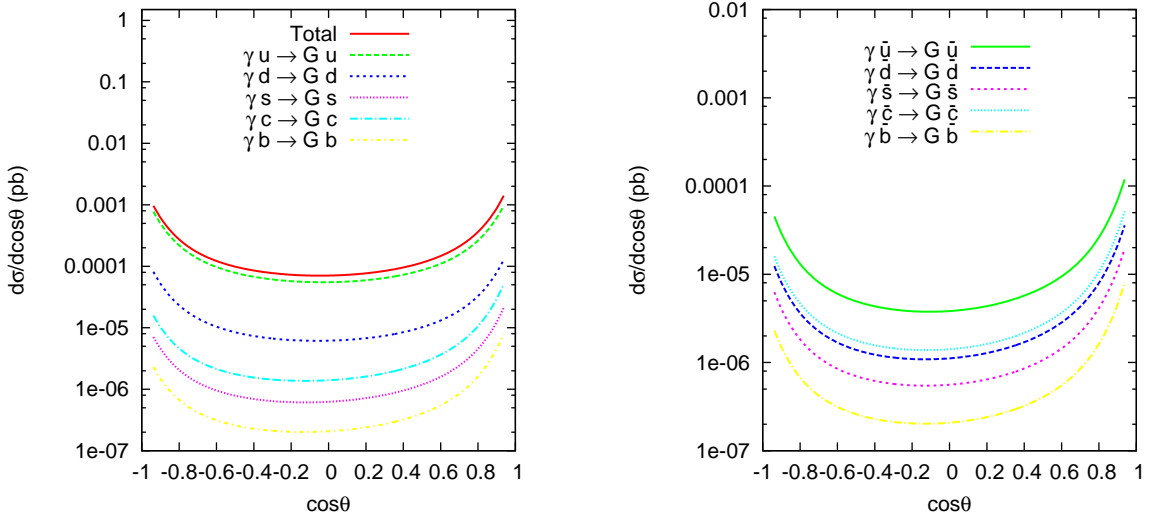}
\caption{The angular distribution of the graviton in the
center-of-momentum frame of the proton-proton system. $\theta$ is
the angle between the outgoing graviton and the incoming photon
emitting intact proton. RS model parameters are chosen to be
$\beta=0.05$ and $m_G=1$ TeV. In the left panel we present total
cross section (solid line) which represents the sum of all
contributions from subprocesses in (\ref{subprocesses}) and cross
sections including individual contributions from each subprocess
with quarks. In the right panel we present similar plots but for
subprocesses with antiquarks.\label{fig11}}
\end{figure}

\begin{figure}
\includegraphics[scale=1]{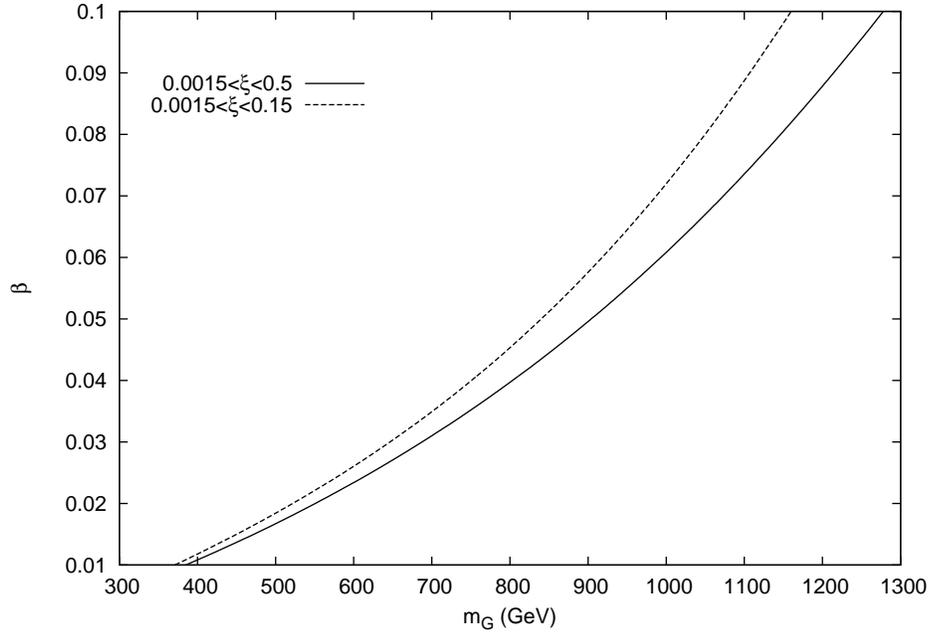}
\caption{The 95\% C.L. limits in the RS parameter space for an
integrated luminosity of 200 $fb^{-1}$ The excluded regions are
defined by the area above the curves. $G \to \gamma \gamma, e \bar
e, \mu \bar \mu$ decay channels with a total branching ratio of
$8\%$ is considered. \label{fig12}}
\end{figure}

In the ADD model of extra dimensions the polarization summed
amplitude square for the subprocess $\gamma q \to G q$ is given by

\begin{eqnarray}
|M|^2=|M_1+M_2+M_3+M_4|^2
\end{eqnarray}

\begin{eqnarray}
|M_{1}|^{2}=-\frac{(qg_e)^2}{4\bar
{M_{Pl}}^2}\frac{[40(m^4-(s-t+u)m^2+su)]}{3m^2}
\end{eqnarray}

\begin{eqnarray}
|M_{2}|^{2}=\frac{(qg_e)^2}{3t^2m^4\bar {M_{Pl}}^2}
&&[-6m^{10}+12tm^8+6((s+u)^2-2t^2)m^6-2t(6s^2+ts-6t^2+ \nonumber \\
&&6u^2+tu)m^4+2t^2(4s^2+(t-4u)s-(3t-4u)(t+u))m^2- \nonumber \\
&&2t^3(s^2+u^2)]
\end{eqnarray}

\begin{eqnarray}
M_{1}^{\dag}M_{2}+M_{2}^{\dag}M_{1}=\frac{(qg_e)^2}{2t\bar {M_{Pl}}^2}\frac{[20(-2m^6+(2s+t+2u)m^4-2t(s+u)m^2+t^3+2stu)]}{3m^2} \nonumber \\
\end{eqnarray}

\begin{eqnarray}
|M_{3}|^{2}=-\frac{(qg_e)^2}{48s^2m^4\bar {M_{Pl}}^2}
&&[4s(2m^8-(5s+t-4u)m^6+2(2s^2-4us+(t+u)(t+2u))m^4 \nonumber \\
&&-(s+3t-2u)(-s+t+u)^2m^2+2st(-s+t+u)^2)]
\end{eqnarray}

\begin{eqnarray}
M_{1}^{\dag}M_{3}+M_{3}^{\dag}M_{1}=-\frac{(qg_e)^2}{24sm^4\bar
{M_{Pl}}^2}
&&[32m^8+(-60s+4t-68u)m^6+8(3s^2+(t+7u)s+ \nonumber \\
&&(t+u)(2t+5u))m^4+4(s^3-3(t-u)s^2+3(t-5u)(t+u)s \nonumber \\
&&-(t+u)^3)m^2-16s(s-t-u)u(t+u)]
\end{eqnarray}

\begin{eqnarray}
M_{2}^{\dag}M_{3}+M_{3}^{\dag}M_{2}=-\frac{(qg_e)^2}{12tsm^4\bar
{M_{Pl}}^2}
&&[2((13s+3t-3u)m^8-2(12s^2+9ts-4t^2+4(s+t)u)m^6 \nonumber \\
&&+(12s^3+(6t+9u)s^2+t(25t+13u)s-(t-u)(13t^2+24ut \nonumber \\
&&+3u^2))m^4-(s^4+(u-6t)s^3+(11t^2+16ut-u^2)s^2+ \nonumber \\
&&(8t^3+15ut^2-2u^2t-u^3)s-2t(t-u)(t+u)(t+6u))m^2 \nonumber \\
&&-2t(s-t-u)(u^3-t^2u+s^2(2t+3u)))]
\end{eqnarray}

\begin{eqnarray}
|M_{4}|^{2}=-\frac{(qg_e)^2}{48u^2m^4\bar {M_{Pl}}^2}
&&[4u(2m^8+(4s-t-5u)m^6+2(2s^2+3ts-4us+t^2+2u^2)m^4 \nonumber \\
&&+(2s-3t-u)(s+t-u)^2m^2+2t(s+t-u)^2u)]
\end{eqnarray}

\begin{eqnarray}
M_{1}^{\dag}M_{4}+M_{4}^{\dag}M_{1}=-\frac{(qg_e)^2}{24um^4\bar
{M_{Pl}}^2}
&&[4(8m^8+(-17s+t-15u)m^6+2(5s^2+7(t+u)s+2t^2 \nonumber \\
&&+3u^2+tu)m^4+(-(s+t)^3-3(5s-t)u(s+t)+u^3+ \nonumber \\
&&3(s-t)u^2)m^2+4s(s+t)(s+t-u)u)]
\end{eqnarray}

\begin{eqnarray}
M_{2}^{\dag}M_{4}+M_{4}^{\dag}M_{2}=-\frac{(qg_e)^2}{12tum^4\bar
{M_{Pl}}^2}
&&[2((-3s+3t+13u)m^8-2(-4t^2+9ut+12u^2+ \nonumber \\
&&4s(t+u))m^6+(3s^3+21ts^2-11t^2s-13t^3+12u^3+ \nonumber \\
&&3(3s+2t)u^2+t(13s+25t)u)m^4-(u^4+(s-6t)u^3- \nonumber \\
&&(s^2-16ts-11t^2)u^2-(s^3+2ts^2-15t^2s-8t^3)u+ \nonumber \\
&&2(s-t)t(s+t)(6s+t))m^2+2t(s+t-u)(s^3-t^2s+ \nonumber \\
&&3u^2s+2tu^2))]
\end{eqnarray}

\begin{eqnarray}
M_{3}^{\dag}M_{4}+M_{4}^{\dag}M_{3}=\frac{(qg_e)^2}{48sum^4\bar
{M_{Pl}}^2}
&&[4(8tm^8+(s^2-15ts-2us-30t^2+u^2-15tu)m^6+ \nonumber \\
&&(-6s^3+6us^2+(26t^2+72ut+6u^2)s-12t^3-6u^3+ \nonumber \\
&&26t^2u)m^4+(3s^4+(13t+8u)s^3-(13t^2+13tu+22u^2)s^2 \nonumber \\
&&-(13t^3+18ut^2+13u^2t-8u^3)s+(2t-3u)(t-u) \nonumber \\
&&(5t+u))m^2+2(s-t-u)(s+t-u)(s^3-us^2-(t+u)^2s\nonumber \\
&&+u^3-t^2u))]
\end{eqnarray}
where $M_1$, $M_2$, $M_3$ and $M_4$  are the amplitudes of the
Feynman diagrams in Fig.\ref{fig2}, $s,t$ and $u$ are Mandelstam
parameters, $m$ is the mass of the individual KK graviton,
$g_e=\sqrt{4\pi\alpha}$, $q=-\frac{1}{3}$ for d,s and b quarks and
$q=+\frac{2}{3}$ for u and c quarks. During amplitude calculations
we neglect the mass of quarks. This is a good approximation at the
LHC energies.

The amplitude square in the RS model can be obtained by the
following replacement: $\bar {M_{Pl}} \to \sqrt 2 \;\Lambda_\pi$

\newpage

\clearpage

\begin{table}
\caption{The 95\% C.L. bounds on $M_D$ for various integrated LHC
luminosities and forward detector acceptance of $0.0015<\xi<0.5$.
Bounds are given in units of GeV. \label{tab1}}
\begin{ruledtabular}
\begin{tabular}{ccccc}
Luminosity&$\delta=3$ &$\delta=4$&$\delta=5$&$\delta=6$ \\
\hline
30$fb^{-1}$ &5314 &4714 &4486 &4400 \\
50$fb^{-1}$ &5886 &5143 &4829 &4686 \\
100$fb^{-1}$ &6743 &5771 &5343 &5114 \\
200$fb^{-1}$ &7771 &6514 &5886 &5571 \\
\end{tabular}
\end{ruledtabular}
\end{table}

\begin{table}
\caption{The 95\% C.L. bounds on $M_D$ for various integrated LHC
luminosities and forward detector acceptance of $0.0015<\xi<0.15$.
Bounds are given in units of GeV. \label{tab2}}
\begin{ruledtabular}
\begin{tabular}{ccccc}
Luminosity&$\delta=3$ &$\delta=4$&$\delta=5$&$\delta=6$ \\
\hline
30$fb^{-1}$ &4560 &4112 &3552 &3356 \\
50$fb^{-1}$ &5028 &4224 &3804 &3608 \\
100$fb^{-1}$ &5784 &4748 &4224 &3916 \\
200$fb^{-1}$ &6636 &5336 &4692 &4280 \\
\end{tabular}
\end{ruledtabular}
\end{table}

\begin{table}
\caption{The 95\% C.L. bounds on $M_D$ for various integrated LHC
luminosities and forward detector acceptance of $0.0015<\xi<0.5$.
Bounds are given in units of GeV. We impose a cut of $M_{Inv}>600$
GeV on the missing invariant mass. \label{tab3}}
\begin{ruledtabular}
\begin{tabular}{ccccc}
Luminosity&$\delta=3$ &$\delta=4$&$\delta=5$&$\delta=6$ \\
\hline
30$fb^{-1}$ &5111 &4695 &4472 &4389 \\
50$fb^{-1}$ &5695 &5112 &4806 &4670 \\
100$fb^{-1}$ &6500 &5722 &5333 &5110 \\
200$fb^{-1}$ &7472 &6444 &5860 &5556 \\
\end{tabular}
\end{ruledtabular}
\end{table}

\begin{table}
\caption{The 95\% C.L. bounds on $M_D$ for various integrated LHC
luminosities and forward detector acceptance of $0.0015<\xi<0.15$.
Bounds are given in units of GeV. We impose a cut of $M_{Inv}>600$
GeV on the missing invariant mass. \label{tab4}}
\begin{ruledtabular}
\begin{tabular}{ccccc}
Luminosity&$\delta=3$ &$\delta=4$&$\delta=5$&$\delta=6$ \\
\hline
30$fb^{-1}$ &4250 &3833 &3540 &3351 \\
50$fb^{-1}$ &4722 &4166 &3790 &3598 \\
100$fb^{-1}$ &5416 &4667 &4187 &3900 \\
200$fb^{-1}$ &6222 &5220 &4639 &4250 \\
\end{tabular}
\end{ruledtabular}
\end{table}

\end{document}